\documentclass[aps,prl,twocolumn,superscriptaddress,showpacs]{revtex4}
\usepackage{graphicx}
\usepackage{dcolumn}
\usepackage{color}

\begin{document}
\title{{\large Measurement of Deeply Virtual Compton Scattering with a Polarized Proton Target}}
%%%%%%%%%%%% Institutes Number and defintions %%%%%%%%%%%
\newcommand*{\FSU}{Florida State University, Tallahassee, Florida 32306}
\affiliation{\FSU}
\newcommand*{\JLAB}{Thomas Jefferson National Accelerator Facility, Newport News, Virginia 23606}
\affiliation{\JLAB}
\newcommand*{\ANL}{Argonne National Laboratory, Argonne, IL, 60439}
\affiliation{\ANL}
\newcommand*{\ASU}{Arizona State University, Tempe, Arizona 85287-1504}
\affiliation{\ASU}
\newcommand*{\UCLA}{University of California at Los Angeles, Los Angeles, California  90095-1547}
\affiliation{\UCLA}
\newcommand*{\CSU}{California State University, Dominguez Hills, Carson, CA 90747}
\affiliation{\CSU}
\newcommand*{\CMU}{Carnegie Mellon University, Pittsburgh, Pennsylvania 15213}
\affiliation{\CMU}
\newcommand*{\CUA}{Catholic University of America, Washington, D.C. 20064}
\affiliation{\CUA}
\newcommand*{\SACLAY}{CEA-Saclay, Service de Physique Nucl\'eaire, F91191 Gif-sur-Yvette, France}
\affiliation{\SACLAY}
\newcommand*{\CNU}{Christopher Newport University, Newport News, Virginia 23606}
\affiliation{\CNU}
\newcommand*{\UCONN}{University of Connecticut, Storrs, Connecticut 06269}
\affiliation{\UCONN}
\newcommand*{\ECOSSEE}{Edinburgh University, Edinburgh EH9 3JZ, United Kingdom}
\affiliation{\ECOSSEE}
\newcommand*{\FIU}{Florida International University, Miami, Florida 33199}
\affiliation{\FIU}
\newcommand*{\GWU}{The George Washington University, Washington, DC 20052}
\affiliation{\GWU}
\newcommand*{\ECOSSEG}{University of Glasgow, Glasgow G12 8QQ, United Kingdom}
\affiliation{\ECOSSEG}
\newcommand*{\ISU}{Idaho State University, Pocatello, Idaho 83209}
\affiliation{\ISU}
\newcommand*{\INFNFR}{INFN, Laboratori Nazionali di Frascati, 00044 Frascati, Italy}
\affiliation{\INFNFR}
\newcommand*{\INFNGE}{INFN, Sezione di Genova, 16146 Genova, Italy}
\affiliation{\INFNGE}
\newcommand*{\ORSAY}{Institut de Physique Nucleaire ORSAY, Orsay, France}
\affiliation{\ORSAY}
\newcommand*{\ITEP}{Institute of Theoretical and Experimental Physics, Moscow, 117259, Russia}
\affiliation{\ITEP}
\newcommand*{\JMU}{James Madison University, Harrisonburg, Virginia 22807}
\affiliation{\JMU}
\newcommand*{\KYUNGPOOK}{Kyungpook National University, Daegu 702-701, Republic of Korea}
\affiliation{\KYUNGPOOK}
\newcommand*{\MIT}{Massachusetts Institute of Technology, Cambridge, Massachusetts  02139-4307}
\affiliation{\MIT}
\newcommand*{\UMASS}{University of Massachusetts, Amherst, Massachusetts  01003}
\affiliation{\UMASS}
\newcommand*{\MOSCOW}{Moscow State University, General Nuclear Physics Institute, 119899 Moscow, Russia}
\affiliation{\MOSCOW}
\newcommand*{\UNH}{University of New Hampshire, Durham, New Hampshire 03824-3568}
\affiliation{\UNH}
\newcommand*{\NSU}{Norfolk State University, Norfolk, Virginia 23504}
\affiliation{\NSU}
\newcommand*{\OHIOU}{Ohio University, Athens, Ohio  45701}
\affiliation{\OHIOU}
\newcommand*{\ODU}{Old Dominion University, Norfolk, Virginia 23529}
\affiliation{\ODU}
\newcommand*{\PITT}{University of Pittsburgh, Pittsburgh, Pennsylvania 15260}
\affiliation{\PITT}
\newcommand*{\RPI}{Rensselaer Polytechnic Institute, Troy, New York 12180-3590}
\affiliation{\RPI}
\newcommand*{\RICE}{Rice University, Houston, Texas 77005-1892}
\affiliation{\RICE}
\newcommand*{\URICH}{University of Richmond, Richmond, Virginia 23173}
\affiliation{\URICH}
\newcommand*{\SCAROLINA}{University of South Carolina, Columbia, South Carolina 29208}
\affiliation{\SCAROLINA}
\newcommand*{\UNIONC}{Union College, Schenectady, NY 12308}
\affiliation{\UNIONC}
\newcommand*{\VT}{Virginia Polytechnic Institute and State University, Blacksburg, Virginia   24061-0435}
\affiliation{\VT}
\newcommand*{\VIRGINIA}{University of Virginia, Charlottesville, Virginia 22901}
\affiliation{\VIRGINIA}
\newcommand*{\WM}{College of William and Mary, Williamsburg, Virginia 23187-8795}
\affiliation{\WM}
\newcommand*{\YEREVAN}{Yerevan Physics Institute, 375036 Yerevan, Armenia}
\affiliation{\YEREVAN}
\newcommand*{\NOWOHIOU}{Ohio University, Athens, Ohio  45701}
\newcommand*{\NOWUNH}{University of New Hampshire, Durham, New Hampshire 03824-3568}
\newcommand*{\NOWMOSCOW}{Moscow State University, General Nuclear Physics Institute, 119899 Moscow, Russia}
\newcommand*{\NOWUMASS}{University of Massachusetts, Amherst, Massachusetts  01003}
\newcommand*{\NOWMIT}{Massachusetts Institute of Technology, Cambridge, Massachusetts  02139-4307}
\newcommand*{\NOWURICH}{University of Richmond, Richmond, Virginia 23173}
\newcommand*{\NOWCUA}{Catholic University of America, Washington, D.C. 20064}
\newcommand*{\NOWECOSSEE}{Edinburgh University, Edinburgh EH9 3JZ, United Kingdom}
\newcommand*{\NOWGEISSEN}{Physikalisches Institut der Universitaet Giessen, 35392 Giessen, Germany}
\author {S.~Chen} 
\affiliation{\FSU}
\author {H.~Avakian} 
\affiliation{\JLAB}
\author {V.D.~Burkert} 
\affiliation{\JLAB}
\author {P.~Eugenio} 
\affiliation{\FSU}
\author {G.~Adams} 
\affiliation{\RPI}
\author {M.~Amarian} 
\affiliation{\ODU}
\author {P.~Ambrozewicz} 
\affiliation{\FIU}
\author {M.~Anghinolfi} 
\affiliation{\INFNGE}
\author {G.~Asryan} 
\affiliation{\YEREVAN}
\author {H.~Bagdasaryan} 
\affiliation{\YEREVAN}
\affiliation{\ODU}
% ...
\collaboration{The CLAS Collaboration}
     \noaffiliation

%%%%%%%%%%%%%%%%%%%%%%%%%%%%%%%%%%% 
% FOLLOWING A GENERIC TAIL FOR TEX
%   Nothing here relates to author list
%   Suggestion for acknowlegements.  Could be deleted.

\date{\today}

\begin{abstract}
 The longitudinal target-spin asymmetry $A_{\rm UL}$ for the exclusive electroproduction of high energy photons was measured for the first time in $e\vec{p} \rightarrow e^{\prime}p\gamma$. The data have been accumulated at JLab with the CLAS spectrometer using 5.7 GeV electrons and a longitudinally polarized ${\rm NH_3}$ target. A significant azimuthal angular dependence was observed, resulting from the interference of the Deeply Virtual Compton Scattering and Bethe-Heitler processes. The amplitude of the $\sin\phi$ moment is $0.252\pm0.042^{stat}\pm0.020^{sys}$. Theoretical calculations are in good agreement with the magnitude and the kinematic dependence of the target-spin asymmetry, which is sensitive to the Generalized Parton Distributions $\widetilde{H}$ and $H$.
\end{abstract}
\maketitle
Generalized parton distributions (GPDs) have in recent years been recognized as a versatile tool to investigate and describe the structure of hadrons at the quark-gluon level. They are closely related to conventional parton distributions and also to hadronic form factors, and contain information that cannot be accessed by either of these quantities. Important aspects where GPDs can provide new insight are the spatial distributions of quarks and gluons within the nucleon and the contribution of quark orbital angular momentum to the nucleon spin. GPDs contain the information needed to construct a multi-dimensional image of the internal structure of the nucleon. The role of GPDs in hard exclusive reactions and their relation to the nucleon's spatial structure and orbital angular momentum are discussed in detail in Refs.~\cite{Mull,xji1,rady,goeke,diehl,radyushkin,burkardt,xji2}.\par
\begin{figure}[t]
\includegraphics[height=.125\textheight]{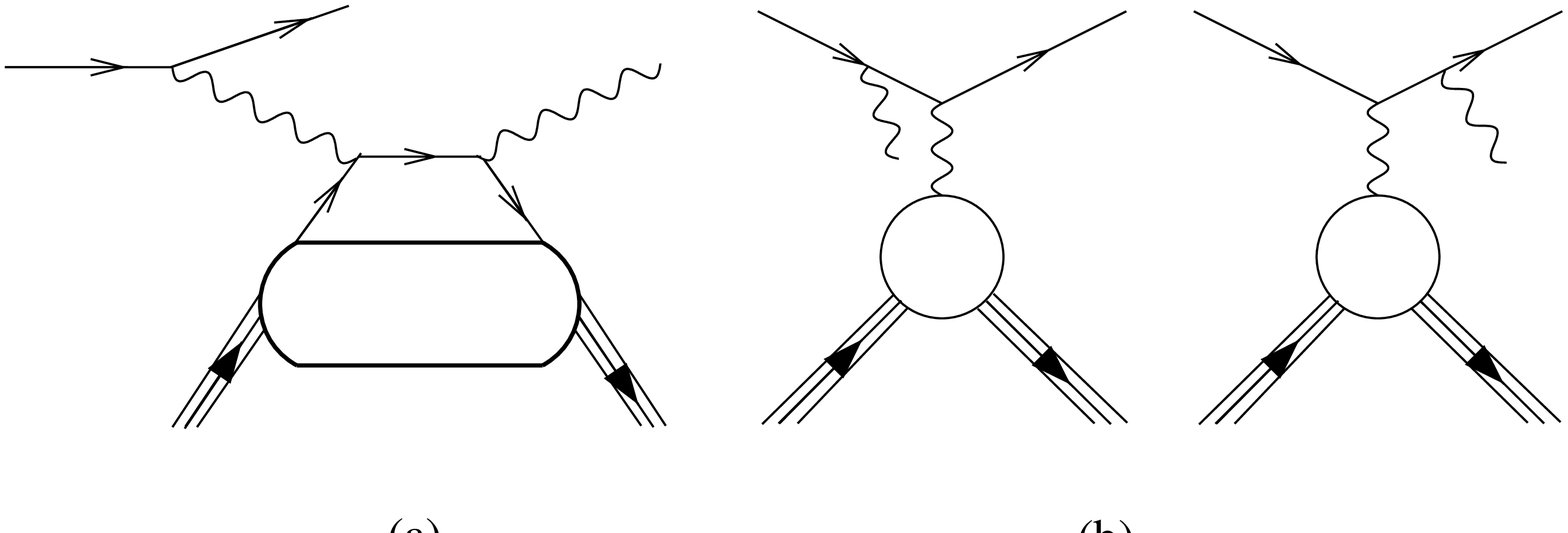}
\caption{Feynman diagrams for DVCS (a) and Bethe-Heitler processes (b) contributing to the amplitude of $ep \to ep\gamma$ scattering.}
\label{fig:dvcs_bh}
\end{figure}

At high photon virtuality $Q^2$ and high energy transfer $\nu$ (Bjorken scaling regime), the scattering amplitude for exclusive processes can be factorized into a hard scattering part (exactly calculable in perturbative QCD), and a nucleon structure part parameterized via GPDs.  This process, called the ``handbag approximation'', is depicted in Fig.~\ref{fig:dvcs_bh}(a) for the case of high-energy photon production. In addition to the dependence on the parton momentum fraction $x$, GPDs depend on two more parameters, the fractional longitudinal momentum transfer $\xi$ to the quark~\cite{GVG1}, and the momentum transfer $t$ to the baryonic system.\par
One of the cleanest processes to access GPDs is Deeply Virtual Compton Scattering (DVCS) in which one quark of the nucleon absorbs a virtual photon producing a real photon with the nucleon left intact. DVCS is most suitable for studying GPDs at moderate energies and in the valence quark regime. At low beam energies, the cross section for DVCS is small and masked by the more copious production of photons from the Bethe-Heitler (BH) process. However, DVCS contributions can be measured directly through the interference of DVCS and BH amplitudes, which result in helicity-dependent cross section differences or asymmetries. The beam spin asymmetry and the target-spin asymmetry can be measured using a polarized electron beam or a polarized target. The two asymmetries are sensitive to different combinations of GPDs and thus provide complementary information~\cite{belitsky}. First experimental results on the beam spin asymmetry $A_{\rm LU}$ with longitudinally polarized beam ($\rm L$) and unpolarized target ($\rm U$) resulting from the DVCS-BH interference have been reported by both the CLAS~\cite{Step} and HERMES~\cite{Aira} collaborations. \par
In this letter we present the first measurement of exclusive DVCS in the target-spin asymmetry measured in the reaction $e \vec{p} \rightarrow e^{\prime} p \gamma$.  The target-spin asymmetry for unpolarized  beam and longitudinally polarized target is defined as
\begin{eqnarray}
\label{eq:AUL0}
A_{\rm UL}(\phi) = \frac{d\sigma^\Uparrow(\phi)-d\sigma^\Downarrow(\phi)}{d\sigma^\Uparrow(\phi)+d\sigma^\Downarrow(\phi)}
\end{eqnarray}
where $\Uparrow$ ($\Downarrow$) represents the target polarization antiparallel (parallel) to the beam direction, and $\phi$ is the angle between the electron scattering plane and the production plane~\cite{trento}. The experiment measures the DVCS contribution through its interference with the Bethe-Heitler (BH) process. In contrast to DVCS, where the photon is emitted from the nucleon, BH photons are emitted from the incoming or scattered electron (Fig.~\ref{fig:dvcs_bh}). While the BH cross section in most of the kinematic region is much larger than the DVCS cross section, the interference of the two contributions enhances the effect of DVCS and produces large cross section asymmetries for the target helicity aligned parallel or antiparallel with the electron beam. In the cross section difference, the helicity-independent BH contribution drops out and only the helicity-dependent interference term remains.\par       
The asymmetry $A_{\rm UL}$ in leading order can be expressed in terms of GPDs~\cite{belitsky}:
\begin{eqnarray}
\label{eq:AUL}
&&A_{\rm UL}(\phi)\propto \{\xi(F_1 + F_2) (\hbox{\bf H}+\frac{\xi}{1+\xi}\hbox{\bf E}) \nonumber\\
&& +F_1 \widetilde{\hbox{\bf H}}- \xi(\frac{\xi}{1+\xi}F_1 + \frac{t}{4M^2}F_2) \widetilde{\hbox{\bf E}} \} \sin\phi,
\end{eqnarray}
where $\widetilde{\hbox{\bf H}}$, $\hbox{\bf H}$, $\widetilde{\hbox{\bf E}}$, and $\hbox{\bf E}$ are sums over quark flavor of the corresponding GPDs with argument $x=\pm \xi$~\cite{belitsky}, $F_1$ and $F_2$ are the known Dirac and Pauli form factors of the proton, and $M$ is the rest mass of the proton. In the range of this experiment the asymmetry is dominated by both $\hbox{\bf H}$ and $\widetilde{\hbox{\bf H}}$, while $\widetilde{\hbox{\bf E}}$ and $\hbox{\bf E}$ are kinematically suppressed. The effect of the $\phi$-dependence of the denominator on the value of the $\sin\phi$ moment extracted from the asymmetry was found negligible. \par 
The data were taken from September 2000 to April 2001 using the CEBAF Large Acceptance Spectrometer (CLAS)~\cite{CLAS}. CLAS is a multi-gap magnetic spectrometer equipped with drift chambers for track reconstruction, scintillator counters for time-of-flight measurements, electromagnetic calorimeters (EC) to identify electrons and photons, and Cherenkov counters (CC) for electron identification. The polar angle coverage of EC is from $8^o$ to $40^o$. Electrons at 5.7~GeV energy were incident on a longitudinally polarized ${\rm ^{15}NH_3}$ target. In this analysis, the asymmetry was averaged over the two beam helicities. The target~\cite{target} polarization was monitored online with a Nuclear Magnetic Resonance (NMR) system, and ranged from 60\% to 80\%. Unpolarized ${\rm ^{4}He}$ and ${\rm ^{12}C}$ targets were used to study the dilution due to the unpolarized material present in the polarized target.\par
The exclusive process $ep\gamma$ was determined by detecting all particles in the final state. Events were selected with the requirements that exactly one positive, one negative, and one neutral track were found for a given trigger, and the particle identifications for these tracks corresponded to an electron, a proton and a photon, respectively. Deep inelastic kinematics was defined by selecting events with $Q^2 > 1\,{\rm GeV^2/c^2}$, $W > 2\, {\rm GeV/c^2}$, and $-t < 0.6\,{\rm GeV^2/c^2}$, where $W$ represents the photon-proton invariant mass.\par
\begin{figure}
\includegraphics[height=.27\textheight]{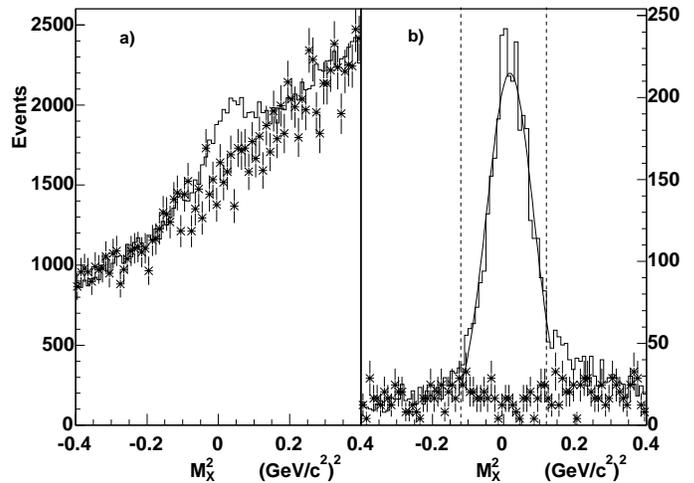}
\caption{Missing mass squared distribution of the system ($ep$) for reaction $e \vec{p} \rightarrow e^{\prime}p\gamma$. Panels (a) and (b) show the $M_{X}^2$ spectrum before and after the cut $\theta_{\gamma X}<2^o$. The $^{12}$C data (stars) are normalized to the $^{15}$NH$_3$ data (solid line) using the negative tail of the $M_{X}^2$. The two dashed lines show the final cut on $M_{X}^2$ to select single photon production.}
 \label{fig:mismass}
\end{figure}

For the ${\rm ^{15}NH_3}$ target, most of the events are from reactions on ${\rm ^{15}N}$ (see Fig.~\ref{fig:mismass}). There is also a large background from $e\vec{p} \rightarrow e^\prime p\pi^0$ events where only one photon from the $\pi^0$ decay was detected. These backgrounds were suppressed by requiring that the detected photon in the over-constrained $e\vec{p} \rightarrow e^\prime p\gamma$ reaction was within 2 degrees of the photon angle predicted from the observed scattered $e^\prime$ and $p$ (see Fig.~\ref{fig:mismass}). The angle cut $\theta_{\gamma X}$ was defined based on Monte Carlo (MC) study. For further analysis events were selected within the missing mass $\vec{p}(e,e^\prime p)X$ range $-0.12 \,{\rm (GeV/c^2)^2} < M_{X}^2 < 0.12\,{\rm (GeV/c^2)^2}$.\par 
\begin{figure}
\includegraphics[height=.25\textheight]{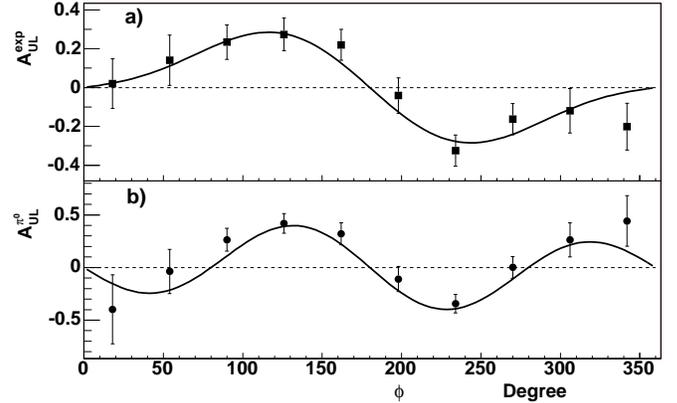}
\caption{Azimuthal angle $\phi$ dependence of the measured target-spin asymmetry for photons (a) and $\pi^0$ (b). The solid curves represent the fitted function $\alpha \sin\phi+\beta \sin2\phi$ with 
parameters $\alpha=0.240\pm0.042$ and $\beta=-0.087\pm0.045$ (a), and  $\alpha=0.109\pm0.056$ and $\beta=-0.319\pm0.061$ (b). In (a), the photon events are still contaminated by $\pi^0$ events.}
\label{fig:SSA1}
\end{figure}

Fig.~\ref{fig:SSA1}(a) shows the azimuthal dependence of $A_{\rm UL}$, which is defined as 
\begin{eqnarray}
A_{\rm UL}(\phi) = \frac{N^\Uparrow(\phi)-N^\Downarrow(\phi)}{f(P_t^\Downarrow N^\Uparrow(\phi)+P_t^\Uparrow N^\Downarrow(\phi))},
\label{eq:AULN}
\end{eqnarray}
where $N^{\Uparrow}$ and $N^{\Downarrow}$ are the luminosity-normalized and acceptance-corrected numbers of $e\vec{p} \to e^\prime p\gamma$ events at positive and negative target helicity respectively, $P_t^\Uparrow$ and $P_t^\Downarrow$ are absolute values of the corresponding target polarizations, and $f=0.901\,\pm\,0.035$ is the dilution factor, which is defined as the ratio of the number of events from hydrogen and from NH$_3$.\par
\begin{figure}
\includegraphics[height=.25\textheight]{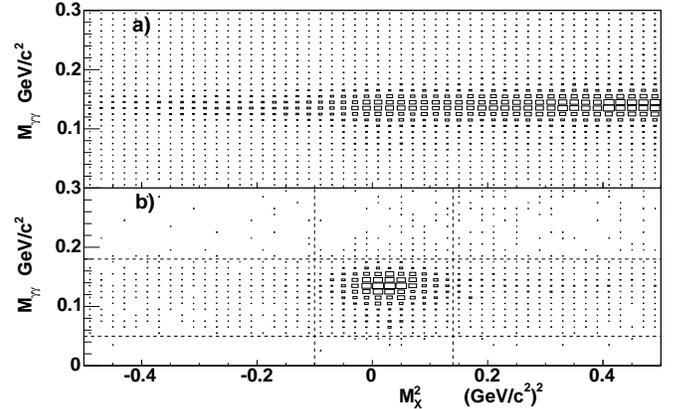}
\caption{The invariant mass of two detected photons vs missing mass squared of the system ($ep$) for reaction $e \vec{p} \rightarrow e^\prime p\gamma \gamma$ before (a) and after (b) the cut $\theta_{\pi^0X}<2.5^o$. Due to the Fermi motion of the protons in $^{15}$N, $M_{X}^2$ for events from nuclear protons is shifted away from $0.018\,{\rm  (GeV/c^2)^2}$. In (b), the four dashed lines show the final cuts on $M_{\gamma\gamma}$ and $M_{X}^2$ respectively.}
 \label{fig:MggvsMMx2}
 \end{figure}

 
The above photon event sample remains contaminated by photons from $\pi^0$ decays that were not removed by the angle cut. In order to correct for this contamination, we analyzed $\pi^0$ events in the same kinematic range as the single $\gamma$ events. Events were selected requiring one electron, one proton, and two detected photons. In Fig.~\ref{fig:MggvsMMx2}(a), a clear band around $M_{\gamma\gamma}=0.135$ ${\rm GeV/c^2}$ shows $\pi^0$ events. Most of these events are from nuclear protons, for which the squared missing mass $M_{X}^2$ is much different from the nominal $M^2_{\pi^0} = 0.018\,{\rm (GeV/c^2)^2}$. Using a similar technique as was used in the DVCS-BH event selection, we placed a cut on the difference of the measured and the predicted $\pi^0$ angles of $\theta_{\pi^0X}<2.5^o$, where the $\pi^0$ angle was reconstructed from measured photons, while the angle of $X$ is predicted using 4-momentum conservation for $e\vec{p} \to e^{\prime} pX$ assuming free proton kinematics. The remaining $\gamma \gamma$ events in Fig.~\ref{fig:MggvsMMx2}(b) cluster around $M_{X}^2= 0.018 \,{\rm (GeV/c^2)^2}$, showing that the events from nuclear protons are largely suppressed. The $\pi^0$ events were selected with cuts $0.05\, {\rm GeV/c} \,< M_{\gamma\gamma}<\,0.18 \,{\rm GeV/c}$ and  $-0.1 \,{\rm (GeV/c^2)^2}\, < M_{X}^2 <\, 0.14 \,{\rm (GeV/c^2)^2}$. For the identified $\pi^0$ events, the dilution factor was $f=0.782\,\pm \,0.036$.\par 
Fig.~\ref{fig:SSA1}(b) shows the azimuthal dependence of $A_{\rm UL}^{\pi^0}$, which was used to correct the DVCS asymmetry for $\pi^0$ contamination.  We note that the asymmetry for $\pi^0$ production has a dominant $\sin2\phi$ dependence while the asymmetry for photon production has a dominant $\sin\phi$ dependence even before $\pi^0$ contributions are fully removed from the single photon sample.  

To estimate the remaining $\pi^0$ contamination in the single $\gamma$ events, a MC study was performed. DVCS-BH and $\pi^0$ events were 
generated and passed through GSIM, the GEANT-based simulation software package of the CLAS spectrometer. The output of GSIM was processed 
using the same procedure as was used for the data. The MC $\pi^0$ spectrum was normalized to the number of $\pi^0$ events observed in the data. Following the 
same procedure as was used in the selection of DVCS-BH events, $\pi^0$ events with only one photon detected were selected to simulate the background from $\pi^0$. The $\phi$ dependence of the $\pi^0$ fraction ($F_{\pi^0}$) is shown in Table~\ref{tab:fpi0}.\par       
\begin{table}[h]
\caption{The $\pi^0$ fraction  and statistical uncertainties in observed single photon events}
 \label{tab:fpi0}
\begin{tabular}{|c|c||c|c|}
\hline
$\phi$ (degree)    &  $F_{\pi^0}\pm \Delta F_{\pi^0}$        & $\phi$ (degree)     &    $F_{\pi^0}\pm\Delta F_{\pi^0}$  \\
\hline
 $0 - 36$            &  $0.106\pm0.010$                         & $180 - 216$          &     $0.373\pm0.022$             \\
\hline
 $36 - 72$           &  $0.117\pm0.009$                         & $216 - 252$          &     $0.313\pm0.019$              \\
\hline
 $72 - 108$          &  $0.242\pm0.018$                         & $252 - 288$          &     $0.216\pm0.015$              \\ 
\hline
 $108 - 144$         &  $0.324\pm0.021$                         & $288 - 324$          &     $0.103\pm0.008$              \\ 
\hline
 $144 - 180$         &  $0.414\pm0.023$                         & $324 - 360$          &     $0.101\pm0.007$              \\
\hline
\end{tabular}
\end{table}
Finally, the fully corrected target-spin asymmetry for single $\gamma$ production was determined using equation:
\begin{eqnarray}
A_{\rm UL}^{exp}(\phi) = F_{\gamma}(\phi)A_{\rm UL}(\phi)+F_{\pi^0}(\phi)A_{\rm UL}^{\pi^0}(\phi),
\label{eq:SSA_correction}
\end{eqnarray}
where $A_{\rm UL}^{exp}$ is the experimentally measured asymmetry with the $\pi^0$ background as shown in Fig.~\ref{fig:SSA1}(a), $A_{\rm UL}^{\pi^0}$ is the target-spin asymmetry for $\pi^0$ as shown in Fig.~\ref{fig:SSA1}(b), and $F_{\gamma}=1-F_{\pi^0}$ is the fraction of DVCS-BH.\par
\begin{figure}
\includegraphics[height=.24\textheight]{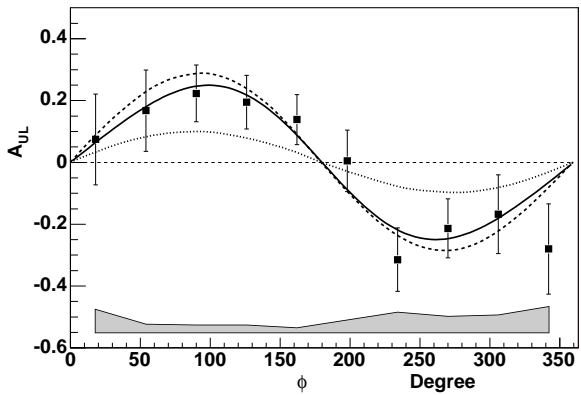}
\caption{The azimuthal angle $\phi$ dependence of the target-spin asymmetry for exclusive electroproduction of photons after subtraction of the $\pi^0$ background. The dashed curve is the full model prediction using the $\xi$-dependent GPD parameterization~\cite{GVG1} ($b_{val}$=$b_{sea}$=1, and ${E}$=$\widetilde {E}$=0) based on MRST02 {unpolarized PDFs~\cite{Martin:2002dr} and polarized PDFs~\cite{Leader:1998qv}} for the twist-2 terms, and higher twists included in those terms. The dotted curve shows the asymmetry when  $\widetilde {H}$=0. The solid curve is described in the text.}
\label{fig:SSA_final}
\end{figure}

The azimuthal dependence of the final asymmetry $A_{\rm UL}$ is shown in Fig.~\ref{fig:SSA_final} at $<Q^2>\,=\,1.82\,{\rm GeV^2/c^2}$, $<-t>\,=\,0.31\,{\rm GeV^2/c^2}$, and $<\xi>\,=\,0.16$. The $\phi$-dependence was fitted with the function $\alpha \sin\phi+\beta \sin2\phi$ (solid curve) with parameters $\alpha=0.252\pm0.042^{stat}\pm0.020^{sys}$, and $\beta = -0.022\pm0.045^{stat}\pm0.021^{sys}$. The $A_{\rm UL}$ is dominated by the $\sin\phi$ term while the $\sin2\phi$ term is compatible with zero within the error bars, indicating that higher twists do not contribute significantly in our kinematical range.\par

\begin{figure}[t]
\includegraphics[width=.5\textwidth]{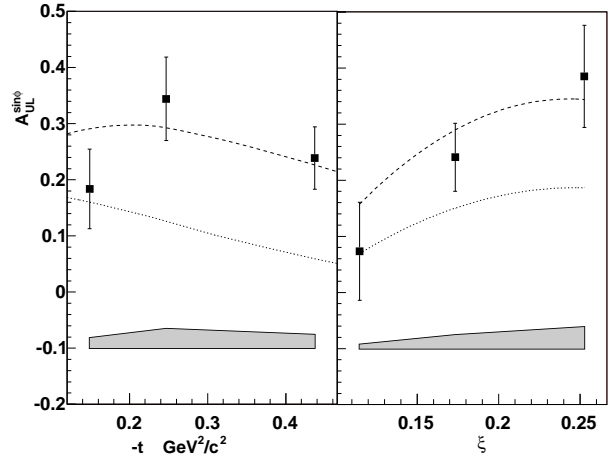}
\caption{The left panel shows the $-t$ dependence of the $\sin\phi$-moment of $A_{\rm UL}$ for exclusive electroproduction of photons, while the right shows the $\xi$ dependence. Curves as in Fig.~\ref{fig:SSA_final}.}
\label{fig:kin_depend}
\end{figure}

To obtain information on the kinematic dependence of the $\sin\phi$-moment of $A_{\rm UL}$ ($A_{\rm UL}^{\sin\phi}$)~\cite{trento}, the data were divided into 3 bins in $\xi$ and $-t$, respectively. The leading term $A_{\rm UL}^{\sin\phi}$ was extracted for each bin. The results are shown in Fig.~\ref{fig:kin_depend}, where the asymmetry was integrated over the other kinematic variables. A clear $\xi$-dependence of $A^{\sin\phi}_{\rm UL}$ is seen, with asymmetries increasing with $\xi$. {The theoretical calculations shown in Fig.~\ref{fig:SSA_final} and Fig.~\ref{fig:kin_depend} have been obtained by including target mass corrections. Unlike Deep Inelastic Scattering (DIS), a full calculation of such corrections is still an open problem for DVCS. We have however included the kinematical higher twist effects in the twist-2 amplitude. In the presence of those effects the GPDs entering in the asymmetry Eq.(~\ref{eq:AUL}) are proportional to GPDs at $(\xi^{\prime}, \xi, t)$, where the difference  between $\xi^{\prime}$ and $\xi$ include terms proportional to  $M^2/Q^2$ and $-t/Q^2$ as shown in Ref.~\cite{GVG1}. As can be noticed on Fig.~\ref{fig:SSA_final} and Fig.~\ref{fig:kin_depend}, the thus obtained theoretical calculation agrees within experimental  uncertainties well with the measurement.}\par

In Fig.~\ref{fig:SSA_final} and Fig.~\ref{fig:kin_depend}, the error bars are statistical, and the systematic uncertainty is shown as a band at the bottom. The sources of systematic uncertainties are identified as the dilution factor calculation ($\sim 4\%$), estimation of target polarization ($\sim 7\%$), $^{15}$N polarization ($\sim 0.5\%$)~\cite{Crabb}, radiative corrections ($<0.1\%$)~\cite{andrei}, evaluation of the $\pi^0$-decay background from MC simulations ($<2.5\%$), and the angle cut ($<5\%$).\par

Combined with the data expected from precision measurements of the beam spin asymmetry which is dominated by GPD $H$~\cite{hep0303006}, these results will allow us to constrain different GPDs. {The target-spin asymmetry in DVCS is also under study at HERMES~\cite{hermesaul}.}\par

In summary, we have presented the target-spin asymmetry for exclusive 
electroproduction of photons. A significant $\sin\phi$ moment of the target-spin asymmetry is observed and is consistent with predictions based on the GPD formalism. The measured asymmetry is consistent with predictions of a large contribution from GPD $\widetilde{H}$. Kinematic dependences of the target-spin asymmetry have also been studied. The leading term $A^{\sin\phi}_{\rm UL}$ increases with increasing $\xi$, in agreement with the model prediction.\par

\begin{acknowledgments}    
We gratefully acknowledge the outstanding efforts of the staff of the Accelerator and Physics Division at Jefferson Lab that made this 
experiment possible. This work was supported in part by the U.S. Department of Energy (DE-FG02-92ER40735)
and the National  Science Foundation,
the Italian Istituto Nazionale di Fisica Nucleare, the
 French Centre National de la Recherche Scientifique,
the French Commissariat \`{a} l'Energie Atomique,
the UK Engineering and Physical Science Research Council, and the Korean Research Foundation.
The Southeastern Universities Research Association (SURA) operates the
Thomas Jefferson National Accelerator Facility for the United States
Department of Energy under contract DE-AC05-84ER40150.
\end{acknowledgments}

\end{document}